\documentclass[twocolumn,prd,floatfix,nofootinbib]{revtex4}
\usepackage{bbm}
\usepackage{amsmath}
\usepackage{graphicx}
\usepackage{dcolumn}
\usepackage{bm}
\usepackage{amssymb}
\usepackage{latexsym}

\def\be{\begin{equation}}
\def\ee{\end{equation}}
\def\ba{\begin{eqnarray}}
\def\ea{\end{eqnarray}}

\usepackage{color}

\begin{document}

\title{Preheating in a DBI Inflation Model}

\author{Jun Zhang$^{a,b}$\footnote{Email: Jun.Zhang@tufts.edu}}
\author{Yifu Cai$^{c}$\footnote{Email: yifucai@physics.mcgill.ca}}
\author{Yun-Song Piao$^{b}$\footnote{Email: yspiao@ucas.ac.cn}}

\affiliation{${}^a$ Department of Physics and Astronomy, Tufts University, Medford, MA 02155, USA}
\affiliation{${}^b$ School of Physics, University of Chinese Academy of Sciences, Beijing 100049, China}
\affiliation{${}^c$ Department of Physics, McGill University, Montr\'eal, QC H3A 2T8, Canada}

\begin{abstract}

We study the preheating process in a model of DBI inflation with a DBI-type inflaton coupling to a canonical entropy field. At the end of inflation, the inflaton field oscillates around its vacuum which can arise from an infrared cutoff parameter on the warp factor and correspondingly the evolution of its fluctuations can be approximately described by a generalized Hill's equation in third order. We study the field fluctuations numerically and show that they could grow exponentially since the instability bands commonly exist in the DBI models if the amplitudes of background oscillations are of order or larger than the cutoff parameter. Our numerical result also reveals that the particle excitation of the matter field is more dramatic than that in usual case since the parametric resonance lasts longer when the effect of a warp factor is taken into account. Therefore, we conclude that the preheating process in the model of DBI inflation could be more efficient than that in standard inflation models.
\end{abstract}

\maketitle

\section{Introduction}

Reheating is a period after inflation during which the vacuum energy of the inßaton field is released into relativistic particles and drive the universe to reach thermal equilibrium \cite{Abbott:1982hn}. Such a process can be realized by introducing a coupling between the inflaton and other matter fields when the inflaton oscillates around its vacuum. Due to these oscillations, it was found that a nonlinear effect of parametric resonance widely exists during the generations of matter field particles, which is so-called {\it preheating} \cite{Traschen:1990sw, Kofman:1994rk, Shtanov:1994ce, Kofman:1997yn}. In recent years, this topic has been extensively studied in the literature within the frame of inflationary cosmology \cite{Boyanovsky:1994me, Baacke:1996se, Cormier:2001iw, Mazumdar:2010sa,Karouby:2011xs}, bubble collision\cite{Zhang:2010qg}, bounce cosmology \cite{Cai:2011ci}, and we refer to Ref. \cite{Allahverdi:2010xz} for a comprehensive review. Accompanied with the process of preheating, it was observed that the field fluctuations could be amplified as well through the parametric resonance \cite{Finelli:1998bu, Finelli:2000ya} or tachyonic process \cite{Abolhasani:2009nb}, although the conversion of these fluctuations upon the curvature perturbation is limited in the case of single field model \cite{Lin:1999ps}.

Most of works addressing on preheating were based on the inflation model with a canonical field. However, the inflation field is not necessarily to be canonical but also can be a general K-essence form\cite{ArmendarizPicon:1999rj}. Due to a non-canonical kinetic term, the propagation of field fluctuations in this type of models is characterized by a sound speed parameter and these fluctuations are frozen not on Hubble radius, but the sound horizon instead. One specific realization of this type of models can be described by a Dirac-Born-Infeld-like (DBI) action\cite{Aharony:1999ti, Myers:1999ps}. Based on brane inflation\cite{Dvali:1998pa}, the model with a single DBI field was initially analyzed in detail in Refs. \cite{Silverstein:2003hf, Alishahiha:2004eh} and later generalized into the multiple throat scenario\cite{Chen:2004gc} and the multiple brane scenario\cite{Cai:2009hw, Cai:2008if}, which has explored a promising window of inflation models without flat potentials. In this model, a warp factor could be applied to provide a speed limit which keeps the inflaton to stay near the top of a potential even if the potential is steep. The phenomenological investigations of a DBI inflation model were extensively studied in the literature. For example, a slow-roll version of multiple DBI scalars were investigated in Refs. \cite{Piao:2002vf, Cline:2005ty}. Such a DBI scalar can also be applied as a curvaton\cite{Li:2008fma} and be embedded into a warped throat\cite{Zhang:2009gw} by taking into account the angular degrees of freedom of the probe brane\cite{Easson:2007dh}, and the behavior of its nonlinear perturbation was studied in \cite{Cai:2010rt}.

An important lesson of the brane inflation model is how to reheat a probe brane and produce Standard Model particles as required by observations in our universe. This process was studied in several detailed stringy backgrounds such as in a D3/D7 system\cite{Brandenberger:2008if} and the KKLMMT scenario\cite{Brandenberger:2007ca}, although these works mainly focus on the leading order contribution of the brane preheating and thus the action of the scalar field still stay canonical. The preheating process after inflation in theories with a non-standard kinetic term was conducted in Refs.
\cite{Lachapelle:2008sy}, in which the perturbation equation exhibits parametric resonance in a modified Hill equation and
becomes more unstable than that described by the Mathieu equation in standard inflationary preheating\cite{Brax:2009hd}. 

In this paper we phenomenologically study the preheating process of a double-field inflation model with a DBI type inflaton coupling to a canonical entropy field. This model can appears in many brane inflation scenarios. Namely, an analytic study of this model based on hybrid inflation was performed in \cite{Davis:2009wg}. The perturbation analysis of this model with trapped branes was studied in \cite{Brax:2011si}. The coupling to other matter fields may lead to the dissipation effect during inflation \cite{Bachlechner:2013fja} and thus could provide a realization of warm brane inflation\cite{Cai:2010wt}. Recently, it was found that the fluctuation of light field may modulate the sound speed of inflaton at very large scale and thus may provide a theoretical origin for dipolar asymmetry as indicated by current CMB observations\cite{Cai:2013gma}. 

The paper is organized as follows. In Section II we briefly introduce the model under consideration and provide the background solution of inflaton after inflation. Then we analyze the preheating process of this model in Section III by performing a detailed numerical computation. Especially, we analyze the parametric resonance effects on both the DBI field fluctuations and the particle creation of the entropy field, respectively. Our result shows that these effects are much more efficient that that happened in a canonical case. Section IV contains the discussion and conclusions. We will work with the reduced Planck mass, $ M_{pl} = 1/  \sqrt{8\pi G}$ where $G$ is the gravitational constant, and adopt the mostly-plus metric sign convention $(-, +, +, +)$. 

\section{Background Dynamics}

To start with, we phenomenologically consider a non-standard model consist of two interacting scalar fields as follows,
\begin{eqnarray}\label{action}
 \mathcal {L} &=&-\frac{1}{f(\phi)} \bigg(\sqrt{1+f(\phi)\partial_{\mu}\phi\partial^{\nu}\phi}-1\bigg) \nonumber\\
 &&
 -V(\phi,\chi)-\frac{1}{2}\partial_{\mu}\chi\partial^{\mu}\chi~,
\end{eqnarray}
where $\phi$ is the inflaton field and the term $f\left(\phi\right)$ is the warp factor motivated by brane inflation models. The $\chi$ field is an entropy field which characterizes the matter fields in the Universe. Its particles are expected to be excited through the interaction between the two fields during preheating and finally to drive the universe to thermal equilibrium. 

One can introduce an infrared cutoff parameter into the warp factor and then the inflaton can obtain a vacuum around $\phi=0$. Specifically, we take the smooth warp factor in form of
\begin{eqnarray}
 f\left(\phi\right)=\frac{\lambda}{\left(\phi^2+\eta^2\right)^2}~,
\end{eqnarray}
where $\eta$ is a parameter of mass dimension and corresponds to the IR cutoff scale. Moreover, as a typical example, we consider the potential of this model to be,
\begin{eqnarray}\label{effectiv potential}
 V\left(\phi,\chi\right) =
 \frac{1}{2}m^2\phi^2+\frac{1}{2}g^2\phi^2\chi^2~,
\end{eqnarray}
in which $m$ is the mass of the inflaton field, and $g$ is the coupling coefficient of the interaction term.

First of all, we examine the background evolution of the inflaton field after inflation by assuming the contribution of $\chi$ field is still negligible within a flat Friedmann-Robertson-Walker universe,
\begin{eqnarray}
 ds^2 = -dt^2 + a^2(t) d\vec{x}^2~.
\end{eqnarray}
This setup is very natural as the initial condition of the preheating process since all other components including the spatial curvature have been diluted out during inflation. By varying the action with respect to the $\phi$ field, one can easily obtain the background equation of motion as follows,
\begin{eqnarray}\label{EOM1}
 \ddot{\phi}+3Hc_s^2\dot{\phi}-\frac{f_{,\phi}}{f^2}+\frac{3}{2}\frac{f_{,\phi}}{f}\dot{\phi}^2+c_s^{3/2} \big(\frac{f_{,\phi}}{f^2}+m^2\phi\big)=0 ~,
\end{eqnarray}
where $c_s$ is the sound speed parameter defined as $c_s=\sqrt{1-f\left(\phi\right)\dot{\phi}^2}$. In Eq. (\ref{EOM1}), the dot denotes the derivative with respect to the cosmic time, and the subscript $_{,\phi}$ denotes the derivative with respect to the scalar $\phi$. In addition, the Einstein field equation yields
the Friedmann equation as follows,
\begin{eqnarray}\label{FRWE1}
 H^2=\frac{8\pi G}{3}\rho_\phi ~,
\end{eqnarray}
and $\rho_\phi$ is the energy density of the inflaton field, which is given by
\begin{eqnarray}\label{energy density}
 \rho_\phi=\frac{1}{f\left(\phi\right)\sqrt{1-f\left(\phi\right)\dot{\phi}^2}}-\frac{1}{f\left(\phi\right)}+\frac{1}{2}m^2\phi^2 ~.
\end{eqnarray}

\begin{figure}
\includegraphics[width=8.5cm]{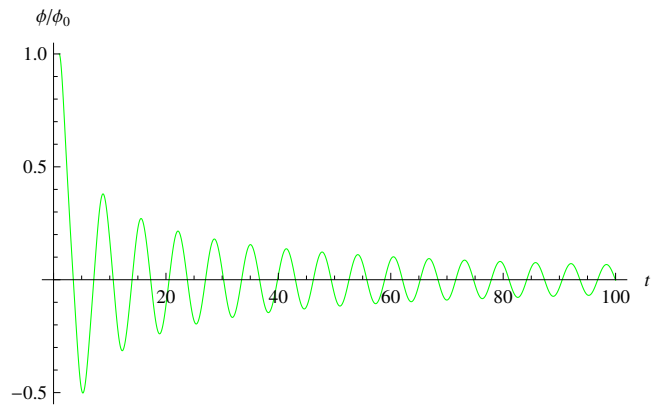}
\caption{\label{fig:sol} Background evolution of the DBI field $\phi$ as a function of the cosmic time $t$ after inflation. $t$
is in the units of $\frac{m}{2\pi}$.}
\end{figure}

To study the dynamics of $\phi$ after inflation, we solve Eq. (\ref{EOM1}) and Eq. (\ref{FRWE1}) numerically and provide the background evolution for $\phi$ in Fig.~\ref{fig:sol}.
From this figure, we can see that $\phi$ is oscillating around the vacuum of its potential after inflation. In the first few
oscillations, the amplitude damps dramatically due to the friction effect brought by the Hubble rate. When the energy density of the universe has dropped down, the friction term becomes small and thus the evolution of $\phi$ can be approximately described as the following solution,
\begin{eqnarray}\label{sol}
 \phi = {\phi_0}~a^{-\frac{3}{2}} \cos\left[m\left(t-t_0\right)\right]+{\rm correction~terms} ~.
\end{eqnarray}
Readers can find more detail in \cite{Bouatta:2010bp}  about these correction terms.

Before the analysis of preheating in this model, we would like to briefly comment on the above background solution. In the usual slow-roll inflation models, the energy density was mainly contributed by the potential energy at the moment when the slow-roll condition was just violated. One should be aware of that for some inflation models the kinetic energy of the inflaton cannot be neglected at the end moment of inflation, namely, the fast roll inflation. Though in DBI inflation the kinetic term is allowed to be large, its contribution is suppressed by the warp factor as shown in Eq. (\ref{energy density}). Therefore, in our case the contribution of the kinetic energy upon preheating is limited and would not spoil the proceeding of preheating. However, it is possible that the kinetic term of inflaton could be important right at the beginning of preheating and may speed up the preheating process. This interesting issue is beyond the scope of the present work and thus we will not address on such a possibility.

\section{Preheating with a DBI field}

After having studied the background evolution of the DBI field at post-inflation stage, in this section we turn to analyze the field fluctuations of $\phi$ and $\chi$. As in usual, we expand these two fields as
\begin{align}
 &\delta\phi\left(\vec{x},t\right) = \tilde{\phi}\left(\vec{x},t\right)-\phi\left(t\right)~,\nonumber\\
 &\delta\chi\left(\vec{x},t\right)= \tilde{\chi}\left(\vec{x},t\right)~,
\end{align}
which can further be expressed in terms of the Fourier modes
\begin{align}
 &\delta\phi\left(x\right) = \int d^3k \big[ a_{\vec{k}} \delta\phi_{k}(t) e^{-ik\cdot x}+a_{\vec{k}}^\dagger \delta\phi_{k}^*(t) e^{ik\cdot x} \big],\nonumber\\
 &\delta \chi \left( x \right) = \int d^3k \big[ b_{\vec{k}}\delta\chi_{k}(t) e^{-ik\cdot x}+b_{\vec{k}}^\dagger\delta\chi_{k}^*(t) e^{ik\cdot x} \big],
\end{align}
where $a_{\vec{k}},b_{\vec{k}}$ and $a_{\vec{k}}^\dagger,b_{\vec{k}}^\dagger$ are the annihilation and creation operators with co-moving momentum $\vec{k}$ for the $\phi$ and $\chi$ fields, respectively. In the following we study the dynamics of field fluctuations by tracking their Fourier modes along the cosmological evolution. 

\subsection{Parametric Resonance of $\delta\phi$}

We first study the evolution of $\delta\phi_k$. For this mode, its dynamics during preheating is associated with the amplitude of the oscillation $\phi_0$. Specifically, we find that the parametric resonance effect is very limit if $\phi_0$ is smaller than the cutoff parameter $\eta$ and thus the situation is the same as the standard preheating phase. However, if $\phi_0$ is larger than $\eta$, we find that the perturbation equation for $\delta\phi_k$ is a generalized Hill's equation in third order and thus allows more instability bands to exist. Therefore, it is much easier to amplify the field fluctuation $\delta\phi$ in a DBI model than that in standard case. 

\subsubsection{A small amplitude}

In the case of $\phi_0 \ll \eta$, the warp factor becomes almost a constant, which is given by $f(\phi) \sim f_0\equiv
\frac{\lambda}{\eta^4}$. Correspondingly, the terms associated with the derivative of $f$ vanished. Then the perturbative equation of $\delta\phi$ in Fourier space becomes
\begin{eqnarray}\label{PEp1}
\ddot{\delta\phi_k}&+&3H\dot{\delta\phi_k} \\ \nonumber
&+& \bigg[c_s^2k^2 +3H\frac{c_s^2-1}{c_s}\frac{\dot{\phi}}{\phi} +\big(4c_s^4-3c_s\big)m^2 \bigg] \delta\phi_k \simeq 0~,
\end{eqnarray}
where we have neglected the interaction with metric perturbation. Note that, for a small amplitude of background oscillation, the damping effect from the friction term is also very limit. Thus, the rest terms are only depending on the sound speed parameter $c_s$. Then we can expand the sound speed as a polynomial function of $f_0\dot{\phi}^2$ around unity and focus on the first order of $f_0\dot{\phi}^2$, since the higher orders red-shift away very quickly. Then, by inserting the background solution (\ref{sol}) into Eq. (\ref{PEp1}) and by performing a variable transformation $m t \rightarrow 2z$, we find the perturbation equation can be reformulated as the Mathieu equation as follow,
%we get
%\begin{eqnarray}
%\ddot{\delta\phi_k}+\left[c_s^2k^2+m^2-\frac{9}{4}f_0m^4\phi_0^2+\frac{9}{4}f_0m^4\phi_0^2\cos{2mt}\right]\delta\phi_k=0.
%\end{eqnarray}
%To compare with the Mathieu equation, we perform the change of
%variables $m_{\sigma}t \rightarrow 2z$, then we get
\begin{eqnarray}\label{ME1}
 \delta\phi_k''+\left[A_k+2q\cos{2z}\right]\delta\phi_k=0,
\end{eqnarray}
where the prime represents for the derivative with respect to $z$, and the coefficients $A_k$ and $q$ are defined as
\begin{eqnarray}
A_k &\equiv& c_s^2\frac{k^2}{m^2}+1-\frac{9}{4}f_0m^2\phi_0^2,\\
q &\equiv& \frac{9}{8}f_0m^2\phi_0^2~.
\end{eqnarray}
The solution of Eq. (\ref{ME1}) has been well studied in the usual preheating theory with a canonical field. This equation reveals that the stability of the solution only depended on $A_k$ and $q$. Specifically, the perturbation modes could experience a process of parametric resonance only when $q \gg 1$ and thus may become unstable against the background. In the meanwhile, however, we observe that, due to the nature of DBI field there is a upper bound on the kinetic term of $\phi$, i.e., $1 - f_0\dot\phi^2 \geq 0$, and thus it yields $q \leq \frac{9}{8}$. Therefore, when $\phi_0$ becomes much smaller than the cutoff parameter $\eta$, the coefficient $q$ is already very small to trigger the occurrence of the parametric resonance. So the perturbation modes of the $\phi$ field are generally stable during preheating.

\subsubsection{A large amplitude}

\begin{figure}
\includegraphics[width=8.5cm]{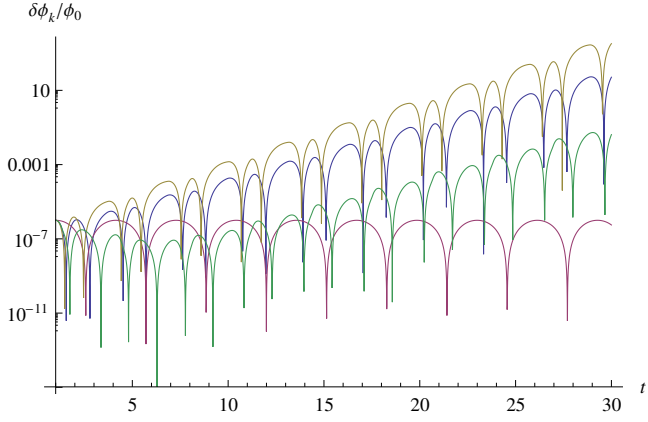}
\caption{\label{fig:dph} 
Evolution of $\delta\phi_k$ with a larger value of $\phi_0$ after inflation. $t$ is in the units of $\frac{m}{2\pi}$. The parameters are given by $\eta=10^{-3}M_{pl}$, $m=10^{-15} M_{pl}$, while $\phi_0=2.5\times10^{-3}M_{pl}, \lambda=10^6$ for blue line, $\phi_0=3\times10^{-3}M_{pl}, \lambda=10^6$ for red line and $\phi_0=2.5\times10^{-3}M_{pl}, \lambda=10^7$ for yellow line. }
\end{figure}

However, when the universe just exits inflation in the DBI model, we have $\phi \sim \eta$. At this moment the dynamics of perturbation modes are very different from the case discussed in previous subsection. If the amplitudes of the background oscillations are large enough, we have to take into account the terms involving $f'(\phi)$. As a result, the perturbation equation of $\phi$ field can be simplified as 
\begin{eqnarray}\label{PEp2}
 \ddot{\delta\phi_k} +\bigg[ c_s^2k^2 +\frac{6\phi^2}{\lambda}F_{c_s} +\big(c_s-\frac{1}{2}c_s^3\big)m^2 \bigg] \delta\phi_k \simeq 0~,
\end{eqnarray}
where we have neglected the terms include $H$ with the same reason mentioned above, and we have defined $F_{c_s}$ as follows,
\begin{eqnarray}
 F_{c_s}\equiv2c_s^4-2c_s^3-c_s^2+1 ~.
\end{eqnarray}
Again, we insert the background solution (\ref{sol}) into Eq. (\ref{PEp2}), and solve the perturbation equation numerically. We show the evolution of the field fluctuation $\delta\phi_k$ in Fig.~\ref{fig:dph}. From the figure, one can find that $\delta\phi_k$ can increase exponentially under certain parameter choices.

\begin{figure}
\includegraphics[scale=0.5]{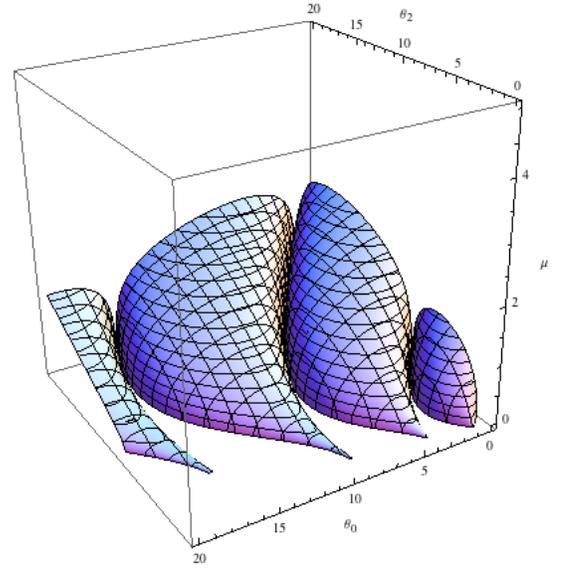}
\caption{The instability band of the Hill's Equation in third order with $\theta_4=0,\theta_6=0$. The numerical simulation is performed by using a 11 $\times$ 11 matrix instead of the infinite one.}
\label{fig:3Hill}
\end{figure}

\begin{figure}
\includegraphics[scale=0.5]{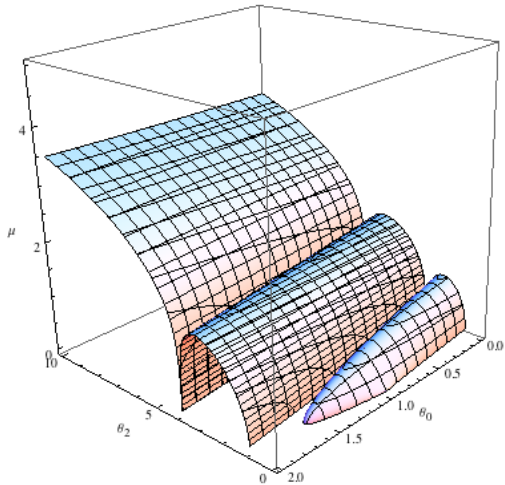}
\caption{The instability band of the Hill's Equation in third order with $\theta_2=0.5\theta_4=-\theta_6$. The numerical simulation is performed by using a 11 $\times$ 11 matrix instead of the infinite one.}
\label{fig:mu}
\end{figure}

Also we perform a semi-analytic calculation by expanding the sound speed parameter $c_s$ up to the first order of $f\dot\phi^2$ and perform the variable transformation $m t \rightarrow 2z$. Therefore, the perturbation equation can be reformulated as a generalized Hill's equation in third order as follows,
\begin{eqnarray}\label{HE}
 \delta\phi_k^{\prime\prime} + \bigg[ \theta_0+\sum_{i=1}^3 \theta_{2i}\cos{2iz} \bigg] \delta\phi_k=0 ~.
\end{eqnarray}
In this perturbation equation we have introduced the following parameters
\begin{align}
 & \theta_0 = \left(1-\frac{A}{B^4}\right)\frac{k^2}{m^2}+1+\frac{9}{128}\frac{A^2}{B^8}+\frac{15}{32}\frac{A}{B^8}~, \\
 & \theta_2 = \frac{k^2}{4m^2}\frac{A}{B^4}-\frac{3}{64}\frac{A^2}{B^8}-\frac{15}{128}\frac{A}{B^8}~, \\
 & \theta_4 = \frac{3}{256}\frac{A^2}{B^8}-\frac{15}{64}\frac{A}{B^8}~, \\
 & \theta_6 = \frac{15}{128}\frac{A}{B^2}~, 
\end{align}
with the coefficients $A \equiv \frac{\lambda m^2}{\phi_0^2}$ and $B \equiv \eta/\phi_0$. The Floquet theorem implies that
the solution of Eq. (\ref{HE}) can be written in the form
\begin{eqnarray}\label{soldph}
 \delta\phi_k = \alpha e^{\mu z}P\left(z\right).
\end{eqnarray}
where $P\left(z\right)$ is periodic in $z$ with a period of $\pi$ and the coefficient $\mu$ is the so-called Floquet exponent which is determined by requiring the vanishment of a determinant of an infinite coefficient matrix.
We perform a numerical simulation by using a $11\times11$ matrix to study the Floquet exponent. The corresponding results of instability bands of the Hill's equation in third order under different parameter choices are shown in Figs. \ref{fig:3Hill} and \ref{fig:mu}, respectively. In Fig.~\ref{fig:3Hill} we can find that, Eq. (\ref{HE}) recovers the usual Mathieu
equation when $\theta_4$ and $\theta_8$ are vanishing. The numerical result is also consistent with the instability bands of Mathieu equation and thus demonstrates the accuracy of our numerical procedure. 

To be more specific, we can describe the evolution of $\delta\phi$ during the post-inflation stage as follows,
\begin{eqnarray}
 \delta\phi_k \simeq \delta\phi_{0k} e^{\mu(\theta)z}.
\end{eqnarray}
Then we can find that there exists a exponential instability if $\mu$ is a real number. Consider the bound of speed limit, we can have $A \sim f\dot\phi^2$ smaller than unity. And since we are interested in $\phi_0 > \eta$, the parameters appeared in the Hill's equation can be approximately expressed as
\begin{align}
 &\theta_0 \simeq \left(1-\frac{A}{B^4}\right)\frac{k^2}{m^2}+1+\frac{15}{32}\frac{A}{B^8}~, \\
 &\theta_2 \simeq \frac{k^2}{4m^2}\frac{A}{B^4}-\frac{15}{128}\frac{A}{B^8}~, \\
 &\theta_4 \simeq -\frac{15}{64}\frac{A}{B^8}~, \\
 &\theta_6 \simeq \frac{15}{128}\frac{A}{B^8}~. 
\end{align}
Thus in the limit of long wavelength we can have $\theta_4 \simeq 2\theta_2$ and $\theta_6 \simeq -\theta_2$. In this case, we again perform the numerical computation on the instability bands of the perturbation equation and show the result in Fig.~\ref{fig:mu}. From this figure, one can obviously find the situation is very different from the case of $\phi_0 \ll \eta$. When $\phi_0$ is larger than $\eta$, the parameter space allowing for the instability bands is quite generic and thus the parametric resonance which can lead to the exponential amplification on the field fluctuations of $\phi$ can be much easily triggered in the period of preheating. 

\subsection{Parametric Resonance of $\delta\chi$}

In this subsection, we continue to study the evolution of $\delta\chi$ after inflation. In order for convenience, we would like to redefine a new field variable,
\begin{eqnarray}
X_k = a^{\frac{3}{2}}\delta\chi_k,
\end{eqnarray}
and thus can get rid of the friction term in the equation of motion. Therefore, the perturbation equation for the $\chi$ field can be simplified as
\begin{eqnarray}\label{PEc1}
 \ddot{X_{k}}+\omega_{X_k}^2X_k = 0~,
\end{eqnarray}
where the effective frequency $\omega_{X_k}$ is define as
\begin{eqnarray}\label{wc1}
 \omega_{X_k}^2 \equiv \frac{k^2}{a^2}+\frac{3}{4}H^2-\frac{3}{2}\dot{H}+g^2\phi^2 ~.
\end{eqnarray}

The same as the analysis in previous subsection, we perform the variable transformation $ mt \rightarrow 2z$ and then Eq.(\ref{PEc1}) can be rewritten in the form of the Mathieu equation as follows,
\begin{eqnarray}\label{PEc2}
 X_k''+\left[A_{Xk}+2q_{X}\cos{2z}\right]X_k = 0 ~,
\end{eqnarray}
with the parameters
\begin{align}
 &A_{Xk} = \frac{k^2}{a^2m^2}+\frac{\phi_0^2g^2}{2m^2a^3},\\
 &q_X = \frac{\phi_0^2g^2}{4m^2a^3}.
\end{align}
As we mentioned before, the solution of Eq. (\ref{PEc2}) mainly depends on $A_{Xk}$ and $q_X$. If $q_X$ starts with a value $q_X \gg 1$, there exists a period of broad resonance on the excitation of $\chi$ particles. Along with the growth of the scale factor, $q_X$ gradually decreases and then the parametric resonance becomes narrow and eventually will stop. 

\begin{figure}
\includegraphics[width=8.cm]{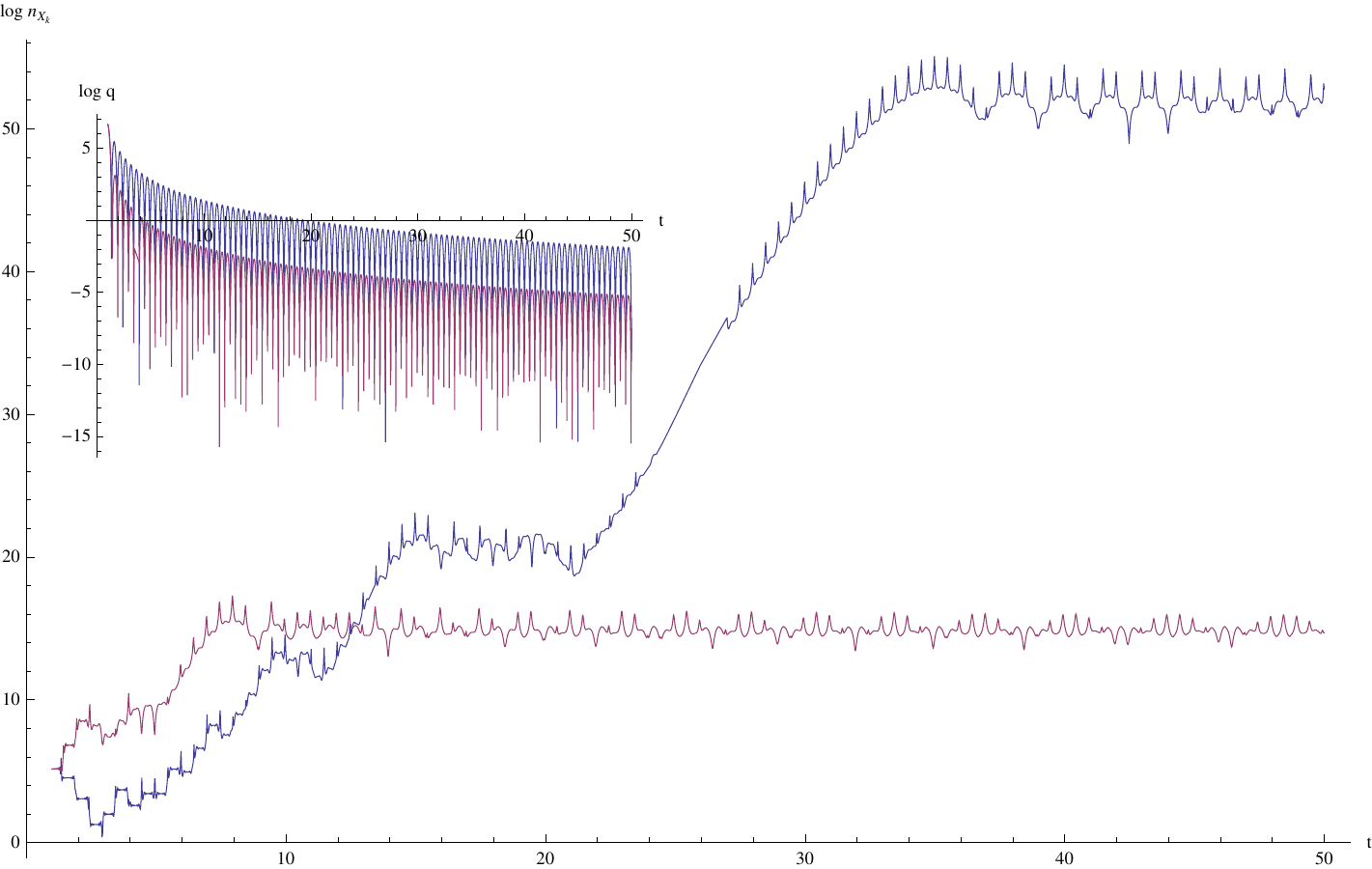}
\caption{\label{fig:compare} 
Particle production of the $\chi$ field in DBI-preheating and that in standard preheating process. The blue line is the numerical solution of DBI-preheating, while the red line is that of the standard one. The cosmic time $t$ is in the unit of $\frac{m}{2\pi}$, and the parameters are given by $m = 7.5 \times 10^{-4}M_{pl},~f_0=5.2\times10^{4}M_{pl}^{-4},~\phi_0=0.07M_{pl},~g=0.55,$ and $k_0=m$.}
\end{figure}

Consider the speed limit again, the condition for the broad resonance to occur can be revised as $ \lambda g^2 \gg 1$ in the case of the DBI model. Therefore, we can see that there always exists the parameter space for the occurrence of the parametric resonance. Also this model can easily give rise to sufficient particle creation for the matter field $\chi$ since the coupling coefficient is allowed to be large due to the existence of warp factor. The production of the $\chi$ particle can be described by the co-moving occupation number, which is defined as
\begin{eqnarray}
 n_k \equiv \frac{\omega_{X_k}}{2}\left(\frac{|\dot{X_k}|^2}{\omega_{X_k}^2}+|X_k^2|\right)-\frac{1}{2} ~.
\end{eqnarray}
Note that Eq. (\ref{PEc2}) is the same as the perturbation equation of the usual preheating theory with standard kinetic terms, but only different on the background evolution. Thus, it is easy to perform a numerical computation to calculate the particle production of the $\chi$ field in our case. To be specific, we show the numerical result in Fig.~\ref{fig:compare} and compare it with the result of the standard preheating theory.

From the above figure, we can find that, the process of DBI-preheating is much more efficient than that of the standard one under the same initial conditions. This is because in Eq. (\ref{EOM1}) the existence of the DBI-terms contribute the corrections to the background evolution which slow down the damping effect of the amplitudes of the inflaton's oscillations. Correspondingly, the decreasing of the index $q_X$ is also much slower than that in the standard case. To show the above statement much clear, we also show the evolutions of the index $q_X$ in Fig.~\ref{fig:compare}. One can easily find the damping process of $q_X$ is much more dramatic in usual case than that in the case of DBI-preheating.

\section{Conclusions}

%Recently, more and more inflation models is studied in the context of string theory. Most of these models involve an DBI-formed effective actions. Compare to the inflation models with a standard kinetic terms, these models often predicts a distinct non-Gaussian primordial spectrum which is favored by the current observational data. However, the preheating of the fields with these DBI-formed action is still unclear.

In this paper, we have studied the preheating of a DBI scalar field $\phi$ which coupling with a standard scalar field $\chi$
after inflation. Under the help of a infrared cut-off on the warp factor, the inflaton field $\phi$ is able to enter a period of oscillation smoothly after inflation. Thus the coupling between the inflaton and other matter fields could provide a suitable environment for the occurrence of preheating. We consider the matter field to be a regular canonical scalar field and the coupling is the same as what people usually consider in standard inflation models. Then we are able to solve the perturbation equations for two fields by tracking their Fourier modes with a fixed co-moving wave number, respectively. We find that, for $\delta\phi$ the dynamics during preheating very much rely on the amplitudes of background oscillations. Specifically, if the amplitude is much smaller than the cutoff parameter of the warp factor, then the resonance parameter is too small to drive the parametric resonance and thus the evolution of the inflaton fluctuation is stable during preheating; however, if the amplitudes of oscillations are of order or larger than the cutoff parameter, the perturbation equation takes the form of a generalized Hill's equation in third order and correspondingly the instability bands in the parameter space are more common than those in canonical models. In this regard, we numerically show the solution to the inflaton's fluctuation would easily encounter a period of exponential growth during preheating and thus could become harmful to the DBI model. Furthermore, we study the dynamics of the matter field fluctuations and show the corresponding particle production is much more efficient than that in usual case since the resonance parameter decreases much slower in the DBI model due to the existence of the warp factor. 

Though the present work only focus on a specific DBI model, our numerical result should give some insight to the preheating theory of a generic inflation model with a non-canonical kinetic term. Especially, our work provides an exact numerical examination to the semi-analytic study performed in \cite{Davis:2009wg}. This work can be viewed as a generalization of the usual preheating theory. 

{\bf Note added}: While this work was being finalized, we noticed a related work \cite{Child:2013ria} by H. Child, J. Giblin, R. Ribeiro and D. Seery on arXiv. Part of our content overlaps with theirs, and our result is similar as well while our focus is on the bands of instabilities of the phase structure of DBI preheating by numerics. Regarding this point, our conclusion is that the preheating process of a DBI scalar is much more efficient than that of a canonical field. 

\begin{acknowledgments}
The work of JZ is supported in part by the Department of Physics and Astronomy at Tufts University. The work of YSP is supported in part by NSFC under Grant No:11075205, 11222546, in part by the ScientiÞc Research Fund of GUCAS (NO:055101BM03), and in part by National Basic Research Program of China under No:2010CB832804. YC is supported in part by an NSERC Discovery Grant. YC is very grateful to YSP for the hospitality during his visit to the University of Chinese Academy of Sciences while this work was initiated three years ago.
\end{acknowledgments}

\end{document}